\documentclass[12pt,titlepage]{article}
\usepackage{amsmath,amssymb,epsfig}
\numberwithin{equation}{section}
\newcommand{\ie}{i.~e.\ }
\newcommand{\eg}{e.~g.\ }
\newcommand{\Castano}{Casta\~{n}o}

\begin{document}

%% Title Page

\begin{titlepage}

\hfill UW/PT-97/27
\vspace{2cm}
\begin{center}
\huge Naturally Nonminimal Supersymmetry\\
\vspace{2cm}
\large David Wright\\
\vspace{21pt}
\normalsize
\textit{Department of Physics,\\ Box 351560,\\ University of 
Washington,\\ Seattle, Washington 98195}\\
\vspace{21pt}
Email: \texttt{wright@phys.washington.edu}
\end{center}

\vspace{5cm}

\begin{center}
\large Abstract \normalsize
\end{center}
We consider the bounds imposed by naturalness on the masses of
superpartners for arbitrary points in nonminimal supersymmetric
extensions of the standard model and for arbitrary messenger
scales.  We discuss appropriate measures of naturalness and
the status of nonminimal supersymmetry in the light of
recent experimental results.

\end{titlepage}

%% Paper

%\begin{frontmatter}
%\title{Naturally Nonminimal Supersymmetry}
%\author{David Wright}
%\address{Department of Physics, University of Washington}
%\begin{abstract}
%We derive a theory of everything and show that it solves the 
%information loss problem associated with black holes.  It also 
%provides a viable inflation senario, solves the strong CP problem, and 
%explains the structure of the fermion mass matrices in the context of 
%supersymmetric M-theoretic string theory.  Issues of duality are 
%explored.
%\end{abstract}
%\end{frontmatter}

\section{Introduction}

The problem of electroweak scale naturalness provides perhaps
the most important motivation for the consideration of
supersymmetric extensions of the standard model (SSMs).
Softly broken supersymmetry stabilizes the electroweak scale against
quadratically divergent radiative corrections, but not against
logarithmic divergences and finite corrections on the order of the
superpartner masses.  It is thus necessary that 
supersymmetry (SUSY) be effectively restored at a scale not much higher
than the electroweak scale in order to avoid fine-tuning.

Previous studies \cite{BG,AC2} have derived bounds on superpartner masses
by requiring that there be no fine-tuning among the GUT-scale parameters of
the supergravity-inspired minimal model with universal
scalar and gaugino masses.  The authors of \cite{DG} relaxed this
universality assumption slightly, allowing for nonuniversal scalar
masses, but maintaining the assumption of gaugino mass universality 
and the assumption that soft SUSY-breaking parameters are set at
the GUT scale.  In this paper, we consider the degree of fine tuning present
at arbitrarily nonminimal points in the SSM parameter space and make no
assumption regarding the so-called messenger scale at which soft SUSY-breaking
parameters are set.
% rewrite, describe new results and outline paper

\subsection{Fine-tuning defined}

In the context of the SSM, the electroweak scale is determined at tree
level by three mass parameters $( m_{x}^{2}, m_{y}^{2}, m_{z}^{2} )$
according to the equations
\begin{align}
\label{mZ}
  \frac{m_Z^2}{2} &=
  \frac{m_x^2 - m_y^2 \tan^2 \beta}{\tan^2 \beta - 1} \\
\label{tanbeta}
  \sin 2 \beta &= \frac{2 m_z^2}{m_x^2 + m_y^2}
\end{align}
Expressions for these mass parameters in terms of SSM parameters will
be given in section \ref{RGsolve}.
Since the mass of the pseudoscalar Higgs boson
$m_A^2 = m_x^2 + m_y^2$,
we can exchange the three mass parameters for $m_Z$, $m_A$, and
$\tan \beta$.
% according to
%\begin{align}
%\label{m1}
%  m_x^2 &=
%  \frac{m_A^2 - \left( m_A^2 + m_Z^2 \right) \cos 2 \beta}{2} \\
%\label{m2}
%  m_y^2 &=
%  \frac{m_A^2 + \left( m_A^2 + m_Z^2 \right) \cos 2 \beta}{2} \\
%\label{m3} 
%  m_z^2 &= \frac{m_A^2 \sin 2 \beta}{2}
%\end{align}
Each point in $( m_Z,m_A,\tan \beta )$ space represents a possible
Higgs sector of the SSM, and the actual $Z$ boson mass defines a
surface in this space.

Qualitatively, the electroweak scale is fine-tuned if a fractionally
small change in the value of a parameter results in a fractionally large
change in $m_Z$.  The degree of fine-tuning in a parameter $a$ can thus be 
quantified by a sensitivity parameter $\Delta$
\begin{equation}
\label{delta}
  \Delta = \frac{\delta m_Z^2}{m_Z^2} \frac{a}{\delta a}
\end{equation}
which was first introduced by Barbieri and Giudice \cite{BG}.
Bounding the allowed degree of fine-tuning, then, corresponds to
requiring that the sensitivity $\Delta$ of $m_Z$ to any parameter
under consideration be less than some specified value;
traditionally, one has taken $\Delta \lesssim 10$, corresponding
to 10\% fine-tuning, as an acceptable level.  Inserting equations 
(\ref{mZ}-\ref{tanbeta}) into (\ref{delta}), we
obtain the expression
\begin{align}
  \Delta =& \frac{2 a}{\left( \tan^2 \beta - 1 \right) m_Z^2}
  \Bigg[ \frac{\delta m_x^2}{\delta a} -
  \tan^2 \beta \frac{\delta m_y^2}{\delta a}
  \nonumber\\
  &+ \frac{\tan \beta}{\cos 2 \beta}
  \left( 1 + \frac{m_Z^2}{m_A^2} \right)
  \left( \sin 2 \beta \left( \frac{\delta m_x^2}{\delta a} +
  \frac{\delta m_y^2}{\delta a} \right) -
  2 \frac{\delta m_z^2}{\delta a} \right) \Bigg]
\label{natural}
\end{align}
We will impose the requirement $\Delta < 10$ to place limits on the parameters
of the SSM as functions of the messenger scale, $\tan \beta$, and
$m_{A}$.
% un-awkwardize sentence

As pointed out by Dimopoulos and Giudice \cite{DG}, the fact that the
coefficients of $\delta m_x^2$ and $\delta m_z^2$ vanish in the
$\tan \beta \rightarrow \infty$
limit implies that $m_y$ must exhibit a dependence on a parameter
in order for the requirement of a particular degree of naturalness to
impose a globally valid limit on the parameter.
This situation arises because the decoupling limit
\begin{equation}
  m_x \sim m_Z \tan \beta
  \qquad
  m_y \sim m_Z
  \qquad
  m_z \sim m_Z \tan^{1/2} \beta
\end{equation}
is a perfectly natural corner of parameter space with large
$\tan \beta$.  The form of the dependence of $m_x$ and $m_z$ on a
parameter are nonetheless relevant in determining how 
a naturalness bound varies with $\tan \beta$ and $m_A$.

We should emphasize that the degree of fine-tuning present in a
particular realization of the SSM depends upon the choice of
a parameter set which we regard as specifying the theory.
Previous studies have taken as a parameter set the
$\overline{\mbox{DR}}$ soft-SUSY breaking parameters at the GUT scale. 
The messengers which communicate SUSY-breaking to the SM particles 
and their superpartners could in principle, however, exist at any scale.
Whatever the scale at which the SUSY-breaking parameters are set,
their effect on the electroweak scale is determined
by integrating the renormalization group (RG) equations which
describe their flow from the messenger scale $\tilde{m}$ down to the
electroweak scale $m_Z$.\footnote{
In fact, this procedure only yields the contributions to $m_Z$
enhanced by factors of $\ln ( \tilde{m} / m_{Z} )$.  Unenhanced
terms, while present, are not only presumably smaller, but also
scheme-dependent.  We therefore ignore the unenhanced terms, but
note that the the limits derived here may therefore be
considered valid only for messenger scales satisfying
 $\ln ( \tilde{m} / m_Z ) \gg 1$.}

In section \ref{RGsolve}, we integrate the relevant one-loop RG equations
in order to find the dependence of the electroweak scale on the
the $\overline{\mbox{DR}}$ parameters at the messenger scale, independent
of any assumptions of universality.  In section \ref{limits}, we apply
equation (\ref{natural}) to obtain limits these on parameters and, in 
turn, on sparticle masses as functions of the messenger scale.
Finally, in section \ref{cousins}, we consider mass limits on sparticles which
first contribute to $m_Z$ at the two-loop level.  

\subsection{Naturalness vs.\ sensitivity}

The equation of sensitivity with unnaturalness has been
criticized by Anderson and \Castano\ \cite{AC}, who rightly point out that
it is inappropriate in some cases.  For example, a small scale $m = \Lambda
e^{-4 \pi/g^2}$ resulting from dynamical symmetry breaking exhibits
a strong sensitivity to $g$ even for perfectly natural values of
$g \sim 1$.  This motivates them to introduce a more refined
measure of naturalness which, instead of simply reflecting the
sensitivity of $m_Z$ to an underlying parameter, compares that
sensitivity to the average sensitivity over the entire allowed
parameter space.  We do not employ their refined definition for
this study.
We do, however, present in this section some considerations which are, to 
our knowledge, original and which we believe elucidate the relationship
between naturalness and sensitivity and clarify the circumstances 
under which their equation is appropriate.

Suppose an underlying parameter set ${\bf a}$ determines the electroweak
scale according to a function $m({\bf a})$ and we imagine that the 
underlying parameters are distributed according to a probability 
distribution $p({\bf a})$.  Then the probability distribution of
electroweak scales is given by
\begin{equation}
\label{p1}
  p(m') = \int \! d{\bf a} \, p({\bf a}) \,
  \delta \! \left( m' - m({\bf a}) \right)
\end{equation}
and the probability of the electroweak scale being as low or lower than
the measured value $m$ is
\begin{equation}
\label{p2}
  P(m) = \int_{0}^{m} {\kern -0.7em}dm' \, p(m')
\end{equation}
We typically imagine the probability density of parameters $p({\bf 
a})$ to be some flat function over an allowed range.  Given some
$p({\bf a})$, the probability $P(m)$ would seem an appropriate
measure of the naturalness of a low electroweak scale.  

In a multidimensional parameter space like that of the SSM, the
examination of the $P(m)$ implied by equations (\ref{p1}-\ref{p2})
for various $p({\bf a})$  seems a daunting task.  We can nevertheless
glean some insight by considering the simplified case in which $m$ depends
depends only on a single parameter $a$, distributed uniformly between
between $0$ and $a_{\max}$.  Then
\begin{equation}
\label{pt1}
  p(m') = \frac{1}{a_{\max}}
  \left( \frac{\delta m}{\delta a} \right)_{a = m^{-1}(m')}^{-1}
\end{equation}
and, \emph{assuming $\delta m / \delta a$ to be relatively constant over
the allowed values of $a$}, we obtain
\begin{equation}
\label{pt2}
  P(m) = \frac{m}{a_{\max}}
  \left. \frac{\delta a}{\delta m} \right|_{a=a_{\max}}
\end{equation}
which implies $\Delta \sim 1 / P$, \ie sensitivity is the inverse
of naturalness.  In the example of Anderson and \Castano, it is
the strong dependence of $\delta m / \delta g$ on $g$ which violates the
assumption made for our toy example and belies the simple relationship
between $\Delta$ and $P$.  Moreover, the integrations described by
equations (\ref{p1}-\ref{p2}) may be easily carried out for the
Anderson and \Castano's example to find that, for $g$ distributed
uniformly on the interval $0 \le g \le {\cal O}(1)$, a hierarchy of
scales $m/\Lambda \sim 10^{-10}$ is not terribly improbable, while
hierarchy $m/\Lambda \sim 10^{-100}$ is quite improbable.  

We believe that equations (\ref{p1}-\ref{p2}) provide a quantitative
definition of naturalness similar in spirit
to that of Anderson and \Castano, but conceptually simpler, since it
allows naturalness to be defined {\it a priori} without reference to
sensitivity.
Our experience with a toy example has suggested that the qualitative
equation of sensitivity to unnaturalness is justified when the
value of $\delta m / \delta a$ is not strongly dependent on the
location in parameter space ${\bf a}$.  Since such an approximation
is indeed valid for the parameters considered here, we believe our
equation of sensitivity with naturalness to be qualitatively valid
for the case under consideration.

\subsection{The calculation sketched}

Before embarking on this calculation, we outline a didactic
method which yields back-of-the envelope estimates of
the limits set by naturalness and the form of their dependence
on $\Delta$ and the messenger scale $\tilde{m}$.  If a mass
parameter $m$ contributes to a higgs mass at tree level, then
we expect from naturalness that.
\begin{align}
  {\cal O}(1) \, m^2 &\lesssim m_Z^2 \, \Delta \\
  m &\lesssim {\cal O} \left( 300 \, \mbox{GeV} \right)
  \left( \frac{\Delta}{10} \right)^{1/2}
\end{align}

If, on the other hand, a particle with mass $m$ contributes at one-loop
via a coupling $g \sim 1$, we expect
\begin{align}
   {\cal O}(1) \, \frac{g^2}{( 4 \pi )^2} \,
   \ln \left( \frac{\tilde{m}}{m} \right) \,
   m^2 &\lesssim m_Z^2 \, \Delta \\
   m &\lesssim {\cal O}  \left( 650 \, \mbox{GeV} \right)
   \left( \frac{33}{t} \frac{\Delta}{10} \right)^{1/2} 
\end{align}
where $t = \ln ( \tilde{m} / m_Z )$, so that $t \sim 33$ for a GUT 
messenger scale.  As this calculation indicates, unlike the bounds on
parameters entering at tree-level, bounds on the masses of particles
contributing at one loop depend on the messenger scale.

Finally, if the particle contributes at two loops, we expect
\begin{align}
   {\cal O}(1) \frac{g^4}{( 4 \pi )^4} \, 
   \ln \left( \frac{\tilde{m}}{m} \right) \,   
   m^2 &\lesssim m_Z^2 \Delta \\
   m &\lesssim \left( 8000 \, \mbox{GeV} \right)
   \left( \frac{33}{t} \frac{\Delta}{10} \right)^{1/2} 
\end{align}
which is, apparently, a significantly weaker bound.

From these considerations, we expect bounds on mass parameters which
enter into higgs masses at tree level to be messenger scale independent
while those which enter as radiative corrections to depend on the
messenger scale approximately as $\ln^{-1/2} ( \tilde{m} / m_Z )$.
All mass bounds are expected to scale as $\Delta^{1/2}$.  Our work in the
following sections will bear out these qualitative expectations.

\section{One-loop RG equations and their solutions}
\label{RGsolve}

The tree-level relations between the parameters $( m_{x}^{2}, m_{y}^{2},
m_{z}^{2} )$ and the soft supersymmetry breaking Higgs masses and
higgsino mass parameter $\mu$ are:
\begin{align}
  m_x^2 &= m_{h_d}^2 + \mu^2 \\
  m_y^2 &= m_{h_u}^2 + \mu^2 \\
  m_z^2 &= m_{ud}^2
\end{align}
We are thus required to integrate the RG equations \cite{MV} for four mass 
parameters --- $m_{h_d}$, $m_{h_u}$, $\mu$ and $m_{ud}$ --- in order
to determine their dependence on messenger scale parameters.
The particles coupled to the higgses by gauge
interactions or the top Yukawa coupling at one loop are:
gauginos, higgsinos, left- and right-handed stops and left-handed
sbottoms\footnote{For $\tan \beta \gtrsim
m_t / m_b$, the bottom quark Yukawa coupling and potentially even the
tau Yukawa coupling are large enough to require consideration.
Since this represents an extreme corner of parameter space and would
much complicate the RG equations, we do not consider this possibility
here.  We also ignore the effects of soft-SUSY breaking trilinear
couplings, which may be present in principle, but are small in most
models.}.
Since these couplings are large, we expect the
masses of these particles to strongly renormalize the higgs
masses; examination of the relevant one-loop RG equations
bears out this expectation.  Because of their strong
effect on the higgs masses, naturalness places strong
constraints on the masses of these particles; for this
reason, the authors of \cite{DG} referred to these particles as
``brothers of the Higgs''.

We begin with the equations for $m_{h_d}$ and $m_{h_u}$.  The
RG equation governing the evolution of the down-type Higgs' mass 
squared reads  
\begin{equation}
\label{hdrge}
  ( 4 \pi )^2 \frac{d m_{h_d}^2}{dt} = - f_{h_d}
\end{equation}
while the up-type Higgs mass squared evolves according to the
coupled equations
\begin{equation}
\label{hurge}
  ( 4 \pi )^2 \frac{d}{dt}
  \begin{pmatrix} m_{h_u}^2 \\ m_q^2 \\ m_u^2 \end{pmatrix} =
  \lambda^2
  \begin{pmatrix} 6 & 6 & 6 \\ 2 & 2 & 2 \\ 3 & 3 & 3 \end{pmatrix}
  \begin{pmatrix} m_{h_u}^2 \\ m_q^2 \\ m_u^2 \end{pmatrix} -
  \begin{pmatrix} f_{h_u} \\ f_q \\ f_u \end{pmatrix}
\end{equation}
Here the inhomogeneous terms are
\begin{align}
   f_{h_{d}} &= 6 g_2^2 m_2^2 + \frac{6}{5} g_1^2 m_1^2 +
   \frac{3}{5} g_1^2 M^2 \\
   f_{h_{u}} &= 6 g_2^2 m_2^2 + \frac{6}{5} g_1^2 m_1^2 -
   \frac{3}{5} g_1^2 M^2 \\
   f_{q} &= \frac{32}{3} g_3^2 m_3^2 + 6 g_2^2 m_2^2 +
   \frac{2}{15} g_1^2 m_1^2 - \frac{1}{5} g_1^2 M^2 \\
   f_{u} &= \frac{32}{3} g_3^2 m_3^2 + \frac{32}{15} g_1^2 m_1^2 +
   \frac{4}{5} g_1^2 M^2
\end{align} 
The gauge couplings and gaugino masses evolve according to
\begin{equation}
\label{gandm}
  g^2_i(t) = g_0^{2} Z_i(t) 
  \qquad
  m_i(t) = \left( m_i \right)_0 Z_i(t)
\end{equation}
where
\begin{equation}
\label{zdef}
  Z_i(t) = \left[ 1 - \frac{2 B_i g_0^2}{(4 \pi)^2} t \right]^{-1}
  \qquad
  B_{1,2,3} = \left( \frac{33}{5} , 1 , -3 \right)  
\end{equation}
The mass $M$, defined by
\begin{equation}
  M^2 = 2 \, \mbox{tr} \left[ Y m^2 \right] =
  m_{h_u}^2 - m_{h_d}^2 + 
  \mbox{tr} \left[ m_q^2 - m_l^2 - 2 m_u^2 + m_d^2 + m_e^2 \right]
\end{equation}
may be taken to be a RG invariant, since the terms proportional
to gaugino masses vanish in its evolution equation, while the
presumably dominant first and second generation scalar masses
are only weakly renormalized by Yukawa couplings.

Before continuing, let us clarify our notation.  Parameters without
any additional annotation (\eg $g$ or $m$) refer to parameters
evaluated at any scale $t = \ln ( m / m_0 )$.  Parameters
annotated with a twiddle (\eg $\tilde{m}$) refer to the value
of the parameter at the messenger scale $\tilde{t} = \ln ( \tilde{m}
/ m_0 )$, and thus constitute what we will regard as the SSM parameters
$a$ in the context of equations (\ref{delta}) and (\ref{natural}).
Finally, parameters with a nought
(\eg $g_0$) refer to GUT-scale parameters evaluated at $t_0 = 0$.

The equation for the evolution of $m_{h_d}^2$ (\ref{hdrge}) can be
readily integrated
\begin{align}
 m_{h_d}^2 &= \tilde{m}_{h_d}^2  -
 \frac{1}{(4 \pi)^2} \int_{\tilde{t}}^t dx \, f_{h_d} \\
 &= \tilde{m}_{h_d}^2 -
 \frac{3}{2} \frac{Z_2^2-\tilde{Z}_2^2}{\tilde{Z}_2^2} \tilde{m}_2^2 -
 \frac{1}{22} \frac{Z_1^2-\tilde{Z}_1^2}{\tilde{Z}_1^2} \tilde{m}_1^2 -
 \frac{1}{22} \ln \left( \frac{Z_1}{\tilde{Z}_1} \right) M^2
\label{mhdsoln}
\end{align}
The solution of the coupled equations (\ref{hurge}) governing the
evolution of $m_{h_u}$ is more complicated but can also be written in
closed form in a useful approximation outlined in an appendix.

%\begin{eqnarray}
%  \begin{pmatrix} m_{h_u}^2 \\ m_q^2 \\ m_u^2 \end{pmatrix} &=&
%  \frac{1}{6}
%  \begin{pmatrix}
%   3 ( 1 + a ) & -3 ( 1 - a ) & -3 ( 1 - a ) \\
%  - ( 1 - a ) & ( 5 + a ) & - ( 1 - a ) \\
%  -2 ( 1 - a ) & -2 ( 1 - a ) & 2 ( 2 + a ) \end{pmatrix}
%  \begin{pmatrix} m_{h_u}^2 \\ m_q^2 \\ m_u^2 \end{pmatrix}_0 +
%  \\ & &
%  \frac{1}{(4\pi)^2} \int_{0}^{t} dx \,
%  \frac{1}{6} \begin{pmatrix}
%   3 ( 1 + b ) & -3 ( 1 - b ) & -3 ( 1 - b ) \\
%  - ( 1 - b ) & ( 5 + b ) & - ( 1 - b ) \\
%  -2 ( 1 - b ) & -2 ( 1 - b ) & 2 ( 2 + b ) \end{pmatrix}
%  \begin{pmatrix} f_{h_u} \\ f_q \\ f_u \end{pmatrix}  
%\end{eqnarray}

The RG equation governing the evolution of the $\mu$-parameter,
\begin{equation}
  \left( 4 \pi \right)^2 \frac{d \mu}{dt} = g(t)
\end{equation}
may be straightforwardly integrated to yield
\begin{equation}
  \mu = \tilde{\mu} \, e^{\int_{\tilde{t}}^t g / ( 4 \pi )^2}
\end{equation}
Here $g(t)$ is a function of gauge and Yukawa couplings whose
specific form is irrelevant for our purposes.

Finally, the RG equation for the evolution of $m_{ud}^2$,
\begin{equation}
  \left( 4 \pi \right)^2 \frac{d m_{ud}^2}{dt} = g(t) \, m_{ud}^2 +
  \left( 6 g_2^2 m_2 + \frac{6}{5} g_1^2 m_1 \right) \mu
\end{equation}
may be likewise straightforwardly integrated
\begin{equation}
  m_{ud}^2 = \left[ \tilde{m}_{us}^2 +
  \left( 3 \frac{Z_2-\tilde{Z}_2}{\tilde{Z}_2} \tilde{m}_2 +
  \frac{1}{11} \frac{Z_1-\tilde{Z}_1}{\tilde{Z}_1} \tilde{m}_1 
  \right) \tilde{\mu} \right] e^{\int_{\tilde{t}}^t g / ( 4 \pi )^2}
\end{equation}

\section{Limits on brothers' masses}
\label{limits}

Having solved the RG equations for the relevant parameters, we can
proceed to apply equation (\ref{natural}) to derive bounds from
naturalness on the parameters which enter into their solutions.

\subsection{Gluino masses}

\begin{figure}
\begin{center}
\epsfig{file=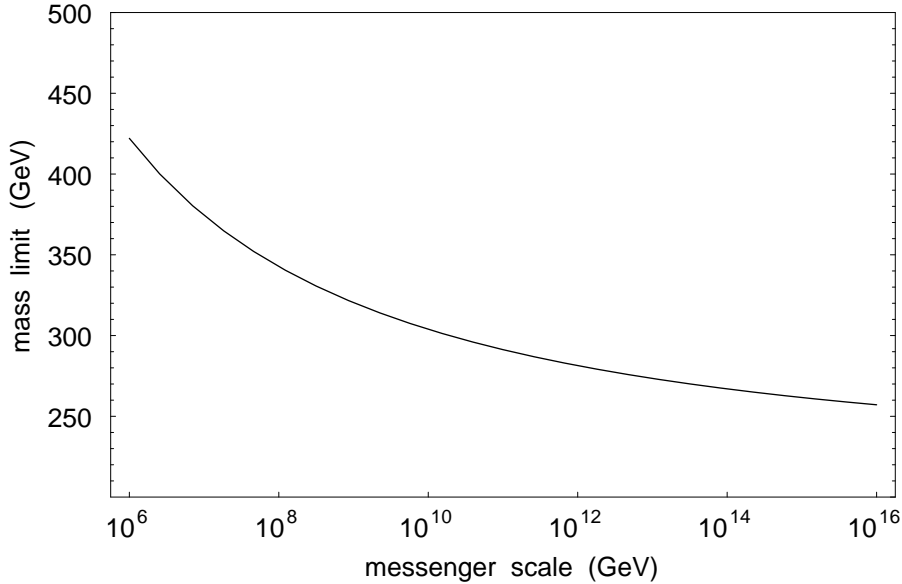}
\caption{$\Delta=10$ limit on gluino mass as a function of the messenger
scale in the large $\tan \beta$ limit.  The scaling of this limit with
$\tan \beta$ and $m_A$ is given by equation (\ref{m3limit}).}
\label{gluino}
\end{center}
\end{figure}

The algebra involved in limiting
gluino mass parameter $m_3$ is easiest, because it enters only into
the expression (\ref{mhusoln}) for $m_{h_u}$.  Applying the naturalness
criterion (\ref{natural}) yields
\begin{multline}
\label{m3limit}
  \frac{32}{9} \frac{m_3^2}{m_Z^2} \frac{1}{Z_3^2}
  \left[ Z_3^2 - \tilde{Z}_3^2 +
  \left( \frac{m_Z}{\tilde{m}} \right)^{\hat{\lambda}} \tilde{W}^{(2)}_3
  - W^{(2)}_3 \right] \\
  < \Delta 
  \left[ 1 - \frac{1}{\tan^2 \beta} \right]
  \left[ 1 + \frac{2 c}{\tan^2 \beta - 1} \right]^{-1}
\end{multline}
where $c = 1 + m_Z^2 / m_A^2 > 1$, and $Z_{i}$, $\hat{\lambda}$ and
$W^{(2)}_{i}$ are defined by equations (\ref{zdef}), 
(\ref{lambdahat}), and (\ref{w2def}), respectively.
We have used equation (\ref{gandm}) to convert a limit on $\tilde{m}_3$
to one on $m_3$ itself.  Note that the limit becomes significantly more
stringent for smaller values of $\tan \beta$.  Figure \ref{gluino} shows the
variation of the limit with the messenger scale $\tilde{m}$.

Because current experiments \cite{D0} already require
$m_3 \gtrsim 200~\mbox{GeV}$
in the context of the minimal model with universal GUT-scale scalar and gaugino
masses, our work allows us to deduce that the gaugino mass in the
MSSM is fine-tuned to at least 11\%; this required degree of 
fine-tuning increases by more than a factor of two as $\tan \beta$ is
reduced to $\tan \beta \sim 2$.  

\subsection{Chargino and neutralino masses}

The analysis of the limits on the bino, wino, and higgsino mass 
parameters $m_1$, $m_2$ and $\mu$ is considerably complicated
by the fact that each of these parameters enters into more than
one RG equation, and in such a way that the application of equation
(\ref{natural}) results in inequalities
involving all three, so that we must solve three inequalities
simultaneously for each value of $\tan \beta$ and $m_A$ in order
to derive bounds on the parameters.

While this procedure can in principle be carried out numerically, the
variation of the resulting bounds with $\tan \beta$, $m_A$, and 
the messenger scale $\tilde{m}$ cannot be displayed in a compact form.
We therefore content ourselves with the presentation of bounds in the
large $\tan \beta$ limit, in which the relevant inequalities
decouple, and with the observation that, as was the case for
gluino masses, mass bounds generically become more stringent as
$\tan \beta$ is lowered.

\begin{figure}
\begin{center}
\epsfig{file=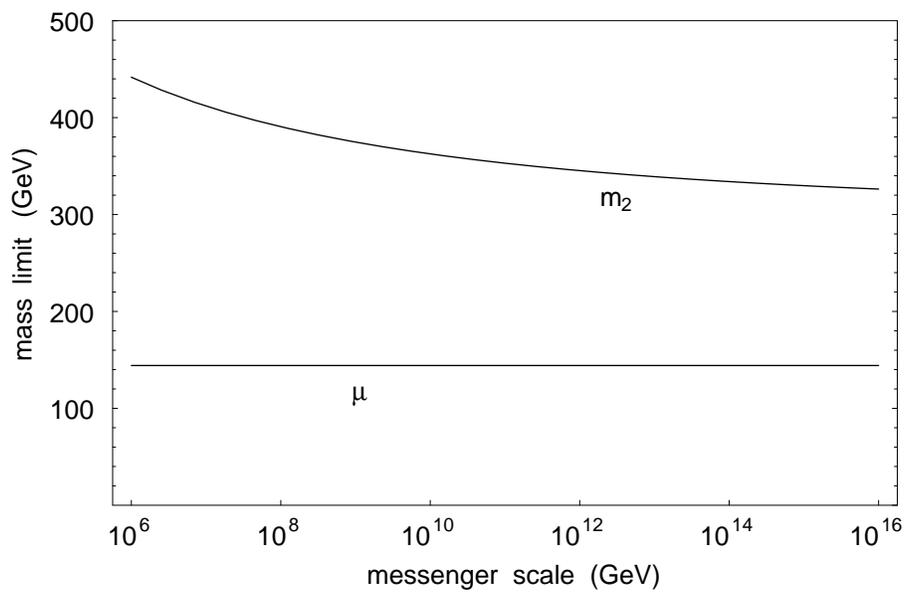}
\caption{$\Delta=10$ limits on wino and higgsino mass parameters as a function
of the messenger scale in the large $\tan \beta$ limit.  The corresponding
limit on the bino mass parameter is greater than 1 TeV.}
\label{winofig}
\end{center}
\end{figure}

In the large $\tan \beta$ limit, as noted previously, only the dependence
of $m_y^2$ on the parameters matters, and the following inequalities are
readily derived:
\begin{align}
\label{m2limit}
  6 \frac{m_2^2}{m_Z^2} \frac{1}{Z_2^2}
  \left[ \left( \frac{m_Z}{\tilde{m}} \right)^{\hat{\lambda}}
  \tilde{W}_{2}^{(2)} - W_2^{(2)} \right] &< \Delta \\
\label{m1limit}
  \frac{8}{99} \frac{m_{1}^{2}}{M_{Z}^{2}} \frac{1}{Z_1^2}
  \left| Z_1^2 - \tilde{Z}_1^2 - \frac{13}{4} \left[
  W^{(2)}_1 - \left( \frac{m_Z}{\tilde{m}} \right)^{\hat{\lambda}}
  \tilde{W}^{(2)}_1 \right] \right| &< \Delta \\
\label{mulimit}
  4 \frac{\mu^2}{m_Z^2} &< \Delta
\end{align}
These limits are displayed in figure \ref{winofig} as a function
of messenger scale. 
Most interesting is the limit on $\mu$, first because it 
is the most stringent and second because, as we expected for 
parameters which enter at tree level, it is independent of
the messenger scale.

Unlike the gluino mass $m_3$, the parameters $m_1$, $m_2$, and $\mu$ 
are not themselves particle masses, but enter into mass matrices which
must be diagonalized in order to determine particle masses.  As
pointed out by Barbieri and Giudice \cite{BG}, however, from the structure
of the relevant mass matrices it follows that the masses of the lightest
chargino and neutralino are constrained by
\begin{align}
  m_{\chi^{\pm}}^2 &< m_W^2 + \min \left( m_2^2 , \mu^2 \right) \\
  m_{\chi^0}^2 &< \min \left( m_1^2 + m_Z^2 \sin^2 \theta_W ,
  m_2^2 + m_Z^2 \cos^2 \theta_W , \mu^2 + \mbox{$\frac{1}{2}$} m_Z^2 \right)
\end{align}
Using these inequalities, our upper bounds on these parameters may be
translated into upper bounds on the masses of the lightest chargino and
neutralino.  In all cases, it is the bound on $\mu$ which dominates,
implying
\begin{align}
  m_{\chi^{\pm}} &< 165 \, \mbox{GeV} \\
  m_{\chi^0} &< 160 \, \mbox{GeV}
\end{align}
Because they arise from the bound on $\mu$, these bounds are 
independent of the messenger scale.  Although they do not
scale exactly as $\Delta^{1/2}$, they do so to a good approximation
for $\Delta \gtrsim 10$.

%Having now set mass limits on all the parameters which determine
%$m_q$ and $m_u$, we may now return to derive limits on these masses.
%\begin{equation}
%  \frac{1}{\Delta} \frac{m_q^2}{m_Z^2} <
%  \frac{1}{6} \frac{ 5 + ( m_Z / \tilde{m} )^{\hat{\lambda}} }{
%  1 - ( m_Z / \tilde{m} )^{\hat{\lambda}} } +
%  \frac{ 1 + ( m_Z / \tilde{m} )^{\hat{\lambda}} }{
%  1 - ( m_Z / \tilde{m} )^{\hat{\lambda}} } +
%\end{equation}

\subsection{Scalar masses}

Now consider the limits on the scalar mass parameters 
$\tilde{m}_{h_u}$, $\tilde{m}_q$, and $\tilde{m}_u$.  From their
contributions to expression (\ref{mhusoln}) for $m_{h_u}$ and
the application of the naturalness criterion (\ref{natural}),
we obtain
\begin{align}
  \frac{\tilde{m}_{h_u}^2}{m_Z^2}
  \left[ 1 + \left( \frac{m_Z}{\tilde{m}} \right)^{\hat{\lambda}} 
  \right] &< \Delta
  \left[ 1 - \frac{1}{\tan^2 \beta} \right]
  \left[ 1 + \frac{2 c}{\tan^2 \beta - 1} \right]^{-1} \\
  \frac{\tilde{m}_{q,u}^2}{m_Z^2}
  \left[ 1 - \left( \frac{m_Z}{\tilde{m}} \right)^{\hat{\lambda}} 
  \right] &< \Delta
  \left[ 1 - \frac{1}{\tan^2 \beta} \right]
  \left[ 1 + \frac{2 c}{\tan^2 \beta - 1} \right]^{-1}
\end{align}

\begin{figure}
\begin{center}
\epsfig{file=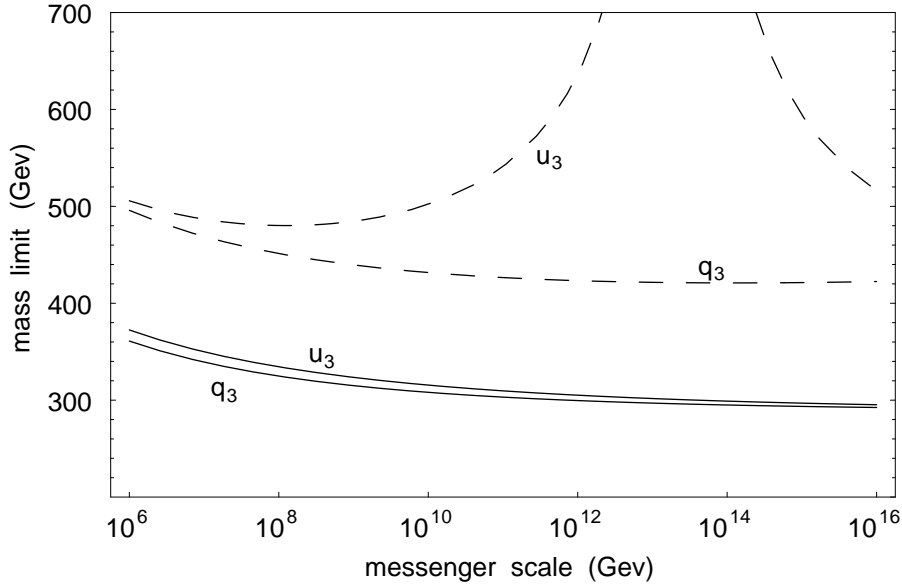}
\end{center}
\caption{$\Delta = 10$ limits on left- and right-handed stop masses in
the large $\tan \beta$ region as a function of the messenger scale.  The
solid lines show the limits implied when scalar masses are large compared
to gaugino masses; when gaugino masses are comparable,
the dashed lines result.  The scaling of the solid lines with $\tan
\beta$ and $m_A$ is given by equations (\ref{q3limit}-\ref{u3limit}).
The bump in the mass limit on right-handed stops near $10^{13}$~GeV is
due to a positive contribution from the bino mass parameter,
which is allowed to be large due to a zero in the coefficient
with which it contributes to $m_{h_u}$ near this scale.}
\label{stop}
\end{figure}

The physical masses of the left- and right-handed stops, as
determined by equations (\ref{mqsoln}) and (\ref{musoln}), receive
contributions from both scalar and gaugino masses.  In the case where
the contributions from scalars dominate, it is easy to translate our
bounds on scalar mass parameters at the messenger scale into bounds on
the stop masses.
\begin{multline}
\label{q3limit}
  m_q^{2} < \frac{m_{Z}^{2} \Delta}{6}
  \left[ 1 - \frac{1}{\tan^2 \beta} \right]
  \left[ 1 + \frac{2 c}{\tan^2 \beta - 1} \right]^{-1}
  \\ \times
  \left\{ \left[ 5 + \left(\frac{m_Z}{\tilde{m}}\right)^{\hat{\lambda}} \right]
  \left[ 1 - \left(\frac{m_Z}{\tilde{m}}\right)^{\hat{\lambda}} \right]^{-1}
  + \left[ 1 - \left(\frac{m_Z}{\tilde{m}}\right)^{\hat{\lambda}} \right]
  \left[ 1 + \left(\frac{m_Z}{\tilde{m}}\right)^{\hat{\lambda}} \right]^{-1} 
  \right\}
\end{multline}
\begin{multline}
\label{u3limit}
  m_u^{2} < \frac{m_{Z}^{2} \Delta}{3}
  \left[ 1 - \frac{1}{\tan^2 \beta} \right]
  \left[ 1 + \frac{2 c}{\tan^2 \beta - 1} \right]^{-1}
  \\ \times
  \left\{ \left[ 2 + \left(\frac{m_Z}{\tilde{m}}\right)^{\hat{\lambda}} \right]
  \left[ 1 - \left(\frac{m_Z}{\tilde{m}}\right)^{\hat{\lambda}} \right]^{-1}
  + \left[ 1 - \left(\frac{m_Z}{\tilde{m}}\right)^{\hat{\lambda}} \right]
  \left[ 1 + \left(\frac{m_Z}{\tilde{m}}\right)^{\hat{\lambda}} \right]^{-1} 
  \right\}
\end{multline}
These bounds are shown as solid lines in figure \ref{stop}.
It is noteworthy that both of these bounds cannot be saturated
simultaneously, since $\tilde{m}_u^2$ contributes negatively to
$m_{q}^{2}$ and vice versa.  The scaling of these limits with 
$\tan \beta$ makes them more stringent by more than a factor of two for
low values of $\tan \beta \sim 2$.

When gaugino masses are comparable to stop masses, their possible
contributions must also be considered.  Inserting the gaugino mass
limits (\ref{m3limit}) and (\ref{m2limit}-\ref{m1limit}) into
equations (\ref{mqsoln}) and (\ref{musoln})  gives the stop mass limits
indicated by the dashed lines in figure \ref{stop}.

\begin{figure}
\begin{center}
\epsfig{file=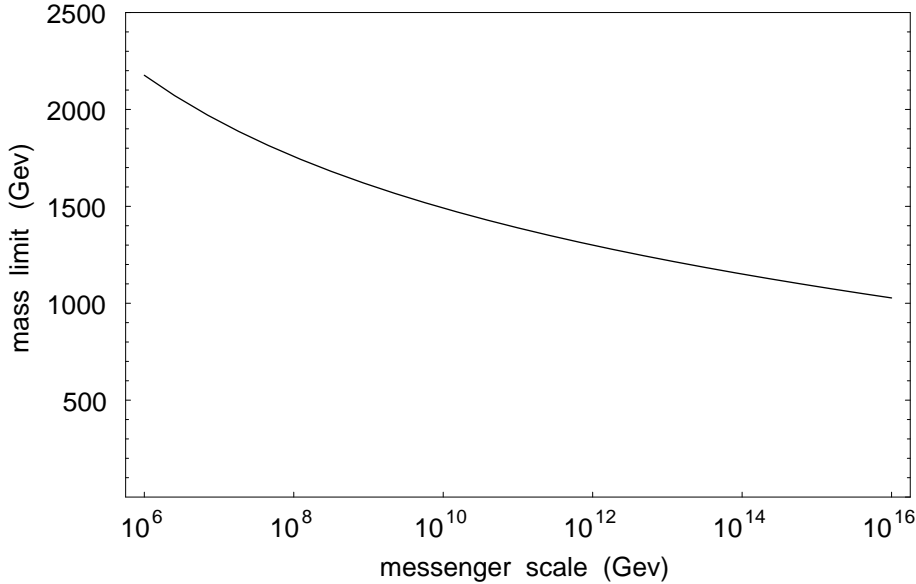}
\end{center}
\caption{$\Delta=10$ limit on generic sparticle masses in $M \ne 0$ models
in the large $\tan \beta$ regime as a function of the messenger scale.
The scaling of this limit with $\tan \beta$ is determined by
equation (\ref{Meqn}).
The limit for right-handed up-type squarks is more stringent by a
factor $\sqrt{2}$.}
\label{oneloop}
\end{figure}

The scalar mass sum $M^2$ enters into the expressions (\ref{mhdsoln},
\ref{mhusoln})  for both $m_{h_d}$ and $m_{h_u}$.
Treating it as a parameter gives a naturalness limit
\begin{equation}
\label{Meqn}
  \frac{1}{11} \frac{M^2}{m_Z^2}
  \ln \left( \frac{\tilde{Z}_1}{Z_1} \right) <
  \Delta \frac{\tan^2 \beta - 1}{\tan^2 \beta + 1 }
\end{equation}
on $M$ and by implication on the masses of the scalars which enter
into it.  The variation of this limit with the messenger scale
$\tilde{m}$ is shown in figure \ref{oneloop} in the large $\tan \beta$
limit; again, the limit becomes more stringent as $\tan \beta$ becomes
smaller.

Of course, $M$ itself should not be regarded as a parameter but as a 
function of the scalar sparticle masses.  As noted by Dimopoulos 
and Giudice~\cite{DG}, if some symmetry, \eg scalar mass degeneracy or
$SU(5)$, ensures $M=0$ then equation (\ref{Meqn}) gives no limit on any
scalar masses.
If, on the other hand, scalar masses are not so constrained, this is an upper
limit on the masses of the scalar partners of fermions of all three
generations.

\section{Cousins of the Higgs}
\label{cousins}

\begin{figure}
\begin{center}
\epsfig{file=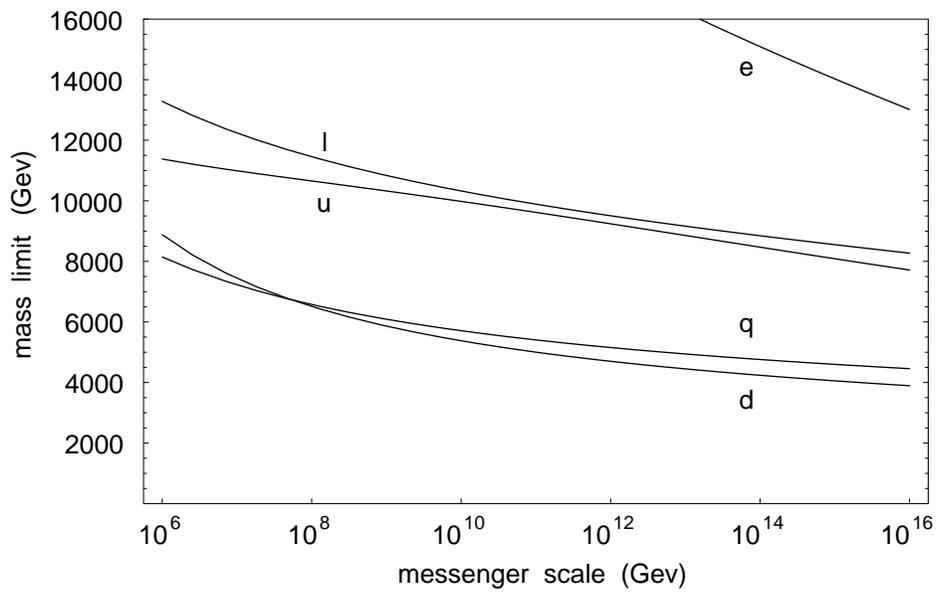}
\end{center}
\caption{$\Delta=10$ limits on sparticle masses from two-loop naturalness
in the large $\tan \beta$ limit as a function of the messenger scale.}  
\label{twoloop}
\end{figure}

If some symmetry principle does protect $M=0$,
the masses of the first and second generation sparticles
enter the one-loop RG equations suppressed by tiny Yukawa couplings
and their masses are thus essentially unlimited by one one-loop radiative
corrections.  On the other hand, at two loops the
inhomogeneous terms in the RG equations described above get
contributions~\cite{MV}
\begin{align}
  ( 4 \pi )^4 \frac{d m_{h_d}^2}{dt} &\supset
  \frac{3}{5} g_1^4 S_1^2 + 3 g_2^4 S_2^2 -
  \frac{6}{5} g_1^2 T^2 \\
  ( 4 \pi )^4 \frac{d m_{h_u}^2}{dt} &\supset
  \frac{3}{5} g_1^4 S_1^2 + 3 g_2^4 S_2^2 +
  \frac{6}{5} g_1^2 T^2 \\
  ( 4 \pi )^4 \frac{d m_q^2}{dt} &\supset
  \frac{1}{15} g_1^4 S_1^2 + 3 g_2^4 S_2^2 + \frac{16}{3} g_3^4 S_3^2 +
  \frac{2}{5} g_1^2 T^2 \\
  ( 4 \pi )^4 \frac{d m_u^2}{dt} &\supset
  \frac{4}{15} g_1^4 S_1^2 + \frac{16}{3} g_3^4 S_3^2 -
  \frac{8}{5} g_1^2 T^2
\end{align}
where
\begin{align}
  S_1^2 &= 2 \, \mbox{$\frac{3}{5}$} \, \mbox{tr} \left[ U(1)_Y^2 m^2 \right]
  \nonumber\\
  &= \mbox{$\frac{1}{5}$} \left( 3 m_{h_u}^2 + 3 m_{h_d}^{2} + \mbox{tr}
  \left[ m_q^2 + 8 m_u^2 + 2 m_d^2 + 3 m_{\ell}^2 + 6 m_e^2 \right] \right) \\
  S_2^2 &= 2 \, \mbox{tr} \left[ SU(2)^2 m^2 \right] =
  m_{h_u}^2 + m_{h_d}^2 + \mbox{tr} \left[ 3 m_q^2 + m_l^2 \right] \\
  S_3^2 &= 2 \, \mbox{tr} \left[ SU(3)^2 m^2 \right] =
  2 m_q^2 + m_u^2 + m_d^2
\end{align}
and
\begin{align}
  T^2 &= g_1^2 T_1^2 + g_2^2 T_2^2 + g_3^2 T_3^2 \\
  T_1^2 &=\mbox{$\frac{3}{10}$} \left( m_{h_u}^2 - m_{h_d}^2 \right) +
  \mbox{tr} \left[
  \mbox{$\frac{1}{30}$} m_q^2 -
  \mbox{$\frac{16}{15}$} m_u^2 +
  \mbox{$\frac{2}{15}$} m_d^2 -
  \mbox{$\frac{3}{10}$} m_l^2 +
  \mbox{$\frac{6}{5}$} m_e^2 \right] \\ 
  T_2^2 &= \mbox{$\frac{3}{2}$} \left( m_{h_u}^2 - m_{h_d}^2 +
  \mbox{tr} \left[ m_q^2 - m_l^2 \right] \right) \\
  T_3^2 &= \mbox{$\frac{8}{3}$} \mbox{tr} \left[ m_q^2 - 2 m_u^2 +
  m_d^2 \right]
\end{align}
Including these mass sums gives a contribution to the up-type higgs
mass
\begin{equation}
\label{mhu2}
  m_{h_u}^2 \supset \frac{\alpha_0}{8 \pi} \left(
  C_{1} S_{1}^{2} + C_{2} S_{2}^{2} + C_{3} S_{3}^{2} + 
  D_{1} T_{1}^{2} + D_{2} T_{2}^{2} + D_{3} T_{3}^{2} \right)
\end{equation}
where
\begin{align}
  C_{1} &= - \frac{2}{99} \left( \tilde{Z}_1 - Z_1\right) -
  \frac{7}{99} \left[ \left( \frac{m_Z}{\tilde{m}}
  \right)^{\hat{\lambda}} \tilde{W}^{(1)}_1 - W^{(1)}_1 \right] \\
  C_{2} &= - 3 \left[ \left( \frac{m_Z}{\tilde{m}}
  \right)^{\hat{\lambda}} \tilde{W}^{(1)}_2 - W^{(1)}_2 \right] \\
  C_{3} &= \frac{8}{9} \left[ 
  \left( \frac{m_Z}{\tilde{m}} \right)^{\hat{\lambda}} \tilde{W}^{(1)}_3
  - W^{(1)}_3 - \tilde{Z}_3 + Z_3 \right]
\end{align}
with
\begin{equation}
\label{w1def}
  W_i^{(1)} = Z_i -
  r_i e^{-r_i Z_i^{-1}} \mbox{Ei} \left( r_i Z_i^{-1} \right)
  \qquad
  r_i = \frac{6 \lambda^2}{B_i g_0^2}
\end{equation}
and
\begin{align}
  D_{1} &= -\frac{2}{11} \left( \tilde{Z}_1 - Z_1 \right) \\
  D_{2} &= \frac{3}{14} \ln \left( \frac{Z_2}{\tilde{Z_2}}
  \frac{\tilde{Z}_1}{Z_1} \right) \\
  D_{3} &= \frac{1}{8} \ln \left( \frac{Z_3}{\tilde{Z_3}}
  \frac{\tilde{Z}_1}{Z_1} \right)
\end{align}
Equation (\ref{mhu2}) implies naturalness limits for all sparticles
of all generations given by
\begin{equation}
  \frac{\alpha_0}{4\pi} \frac{m_s^2}{m_Z^2} \left| C_s \right| < \Delta
\end{equation}
where the coefficients $C_s$ may easily be read off as
\begin{align}
  C_{q} &= \mbox{$\frac{1}{5}$} C_{1} + 3 C_{2} + 2 C_{3} +
  \mbox{$\frac{1}{30}$} D_{1} + \mbox{$\frac{3}{2}$} D_{2} +
  \mbox{$\frac{8}{3}$} D_{3} \\
  C_{u} &= \mbox{$\frac{8}{5}$} C_{1} + C_{3} -
  \mbox{$\frac{16}{15}$} D_{1} - \mbox{$\frac{16}{3}$} D_{3} \\
  C_{d} &= \mbox{$\frac{2}{5}$} C_{1} + C_{3} +
  \mbox{$\frac{2}{15}$} D_{1} + \mbox{$\frac{8}{3}$} D_{3} \\
  C_{l} &= \mbox{$\frac{3}{5}$} C_{1} + C_{2} -
  \mbox{$\frac{3}{10}$} D_{1} - \mbox{$\frac{3}{2}$} D_{3} \\
  C_{e} &= \mbox{$\frac{6}{5}$} C_{1} + \mbox{$\frac{6}{5}$} D_{1}
\end{align}
The resulting limits are shown in figure \ref{twoloop}.  This shows
that it is possible to raise the masses of the q- and d-type
first and second generation squarks to between 4 and 8 TeV, and
those of the other squarks and sleptons even higher, without violating
naturalness.

\section{Conclusions}

We have derived bounds on the masses of
supersymmetric partners from the requirement of naturalness in
arbitrarily nonminimal incarnations of the SSM in which SUSY-breaking
is communicated by messengers at an arbitrary scale.

Examination of the variation of naturalness bounds with messenger scale
reveals that these bounds may be increased from their GUT messenger values
by between 15\% and 100\% by lowering the messenger scale for all 
superpartners except the lightest charginos and neutralinos, for which we
obtain messenger scale independent mass bounds of $\sim 160$~TeV for
10\% fine-tuning.

We have derived the variation of several of these several mass bounds with
$\tan \beta$ and found that naturalness more significantly constrains
sparticle masses in models with low $\tan \beta$ than in models with
high $\tan \beta$.  The mass constraints in the two regimes can differ
by more than a factor of two.

The most problematic constraint for the traditional MSSM with GUT 
scale mediation of SUSY breaking is that placed on the gluino mass
$m_3 \lesssim 260$~GeV.
Present experiments thus require that the MSSM be at least 11\%
fine-tuned and more than 5\% fine-tuned for $\tan \beta \lesssim 2$.
This problem can be evaded by models which significantly lower
the messenger scale or make the gluino the lightest supersymmetric
particle.

As accelerator searches progress, this work will place increasingly
stringent constraints on possible realizations of the SSM.

\section*{Acknowledgments}
The author gratefully acknowledges many
helpful conversations with Ann Nelson.  This work was supported by the
U.S.\ Department of Energy grant number DE-FG03-96ER-40956.

\appendix
\section{Appendix}
Equations (\ref{hurge}) represent an inhomogeneous linear differential
system which may be solved formally in closed form.  In order to
make computational sense of this formal solution, we must examine
the behavior of the top Yukawa coupling $\lambda(t)$, which evolves
according to the RG equation
\begin{equation}
\label{top}
  \frac{d \lambda}{dt} = \frac{\lambda}{(4\pi)^2} \left[
  6 \lambda^2 - \frac{16}{3} g_3^2 - 3 g_2^2 -
  \frac{13}{15}g_1^2 \right]
\end{equation}
This equation lacks a closed-form analytic solution, but an
nonetheless be integrated numerically.  Numerical integration
indicates that the top Yukawa coupling remains within ten percent of
$\lambda \sim 1$ over nearly\footnote{The exception occurs for GUT
scale values of $t$ when $\tan \beta \lesssim 1.5$.  Even an exact
treatment of the evolution of $\lambda$ would not be particularly
helpful in this case, since its behavior in this region becomes highly
sensitive to the exact value of $m_t$.} 
the entire interesting range of $t$ and
$\tan \beta$.  We will therefore perform the integrations considering
$\lambda$ to be constant.  This yields the results
\begin{align}
\label{mhusoln}
  m_{h_u}^{2} =&
  \left[ 1 + \left( \frac{m}{\tilde{m}} \right)^{\hat{\lambda}}
  \right] \frac{\tilde{m}_{h_u}^2}{2} -
  \left[ 1 - \left( \frac{m}{\tilde{m}} \right)^{\hat{\lambda}}
  \right] \frac{\tilde{m}_q^2 + \tilde{m}_u^2}{2}
  \nonumber\\
  &- \frac{8}{9} \frac{Z_3^2 - \tilde{Z}_3^2}{\tilde{Z}_3^2} \tilde{m}_3^2
  - \frac{2}{99} \frac{\tilde{Z}_1^2 - Z_1^2}{\tilde{Z}_1^2} \tilde{m}_1^2
  \nonumber\\
  &+ m_{\lambda}^2 +
  \frac{1}{22} \ln \left( \frac{Z_1}{\tilde{Z}_1} \right) M^2
\end{align}
\begin{align}
\label{mqsoln}
  m_{q}^{2} =&
  \left[ 5 + \left( \frac{m}{\tilde{m}} \right)^{\hat{\lambda}}
  \right] \frac{\tilde{m}_{q}^2}{6} -
  \left[ 1 - \left( \frac{m}{\tilde{m}} \right)^{\hat{\lambda}}
  \right] \frac{\tilde{m}_{h_u}^2 + \tilde{m}_u^2}{6}
  \nonumber\\
  &+ \frac{16}{27} \frac{Z_3^2 - \tilde{Z}_3^2}{\tilde{Z}_3^2} \tilde{m}_3^2
  + \frac{\tilde{Z}_2^2 - Z_2^2}{\tilde{Z}_2^2} \tilde{m}_2^2
  - \frac{5}{297} \frac{\tilde{Z}_1^2 - Z_1^2}{\tilde{Z}_1^2} \tilde{m}_1^2
  \nonumber\\
  &+ \frac{1}{3} m_{\lambda}^2 +
  \frac{1}{66} \ln \left( \frac{Z_1}{\tilde{Z}_1} \right) M^2
\end{align}
\begin{align}
\label{musoln}
  m_{u}^{2} =&
  \left[ 2 + \left( \frac{m}{\tilde{m}} \right)^{\hat{\lambda}}
  \right] \frac{\tilde{m}_{q}^2}{3} -
  \left[ 1 - \left( \frac{m}{\tilde{m}} \right)^{\hat{\lambda}}
  \right] \frac{\tilde{m}_{h_u}^2 + \tilde{m}_q^2}{3}
  \nonumber\\
  &+ \frac{8}{27} \frac{Z_3^2 - \tilde{Z}_3^2}{\tilde{Z}_3^2} \tilde{m}_3^2
  - \frac{\tilde{Z}_2^2 - Z_2^2}{\tilde{Z}_2^2} \tilde{m}_2^2
  + \frac{11}{297} \frac{\tilde{Z}_1^2 - Z_1^2}{\tilde{Z}_1^2} \tilde{m}_1^2
  \nonumber\\
  &+ \frac{2}{3} m_{\lambda}^2 -
  \frac{2}{33} \ln \left( \frac{Z_1}{\tilde{Z}_1} \right) M^2
\end{align}
where
\begin{align}
  m_{\lambda}^2 =&
  \frac{8}{9} \frac{1}{\tilde{Z}_3^2} \left[ W_3^{(2)} -
  \left( \frac{m}{\tilde{m}} \right)^{\hat{\lambda}} \tilde{W}_3^{(2)}
  \right] \tilde{m}_3^2
  + \frac{3}{2} \frac{1}{\tilde{Z}_2^2} \left[ \left( \frac{m}{\tilde{m}} 
  \right)^{\hat{\lambda}} \tilde{W}_2^{(2)} - W_2^{(2)} \right] \tilde{m}_2^2
  \nonumber\\ &+
  \frac{13}{198} \frac{1}{\tilde{Z}_1^2} \left[ \left( \frac{m}{\tilde{m}} 
  \right)^{\hat{\lambda}} \tilde{W}_1^{(2)} - W_1^{(2)} \right] 
  \tilde{m}_1^2 
\end{align}  
with
\begin{equation}
\label{lambdahat}
  \hat{\lambda} = \frac{12 \lambda^2}{(4\pi)^2}
\end{equation}
and
\begin{equation}
\label{w2def}
  W^{(2)}_i = Z_i^2 + r_i Z_i -
  r_i^2 e^{- r_i Z_i^{-1}} \mbox{Ei} \left( r_i Z_i^{-1} \right)
  \qquad
  r_i = \frac{6 \lambda^2}{B_i g_0^2}
\end{equation}

\end{document}